\documentclass[runningheads]{llncs}
\usepackage[T1]{fontenc}
\usepackage{graphicx}
\usepackage{hyperref}
\usepackage{xurl}
\usepackage{amsmath}
\usepackage{booktabs}
\usepackage{colortbl}
\usepackage{cleveref}
\usepackage{multirow}
\usepackage{makecell}
\usepackage[dvipsnames]{xcolor}
\usepackage[multiple,hang,flushmargin]{footmisc}
\usepackage{circledsteps}
\usepackage{adjustbox}
\usepackage{algorithm}
\usepackage{algpseudocode}
\usepackage[numbers]{natbib}
\usepackage{svg}
\usepackage{subcaption} 
\usepackage{comment}
\usepackage{adjustbox}
\usepackage{booktabs}
\usepackage{lipsum}
\usepackage{enumitem}
\usepackage{xspace}
\usepackage{soul}
\usepackage{xcolor}

\renewcommand{\footnotesize}{\fontsize{8pt}{11pt}\selectfont}

\makeatletter
\newcommand{\printfnsymbol}[1]{%
 \textsuperscript{\@fnsymbol{#1}}%
}
\makeatother

\newcommand{\method}{\textsc{hn-moea}\xspace}

\begin{document}
\title{Influence Maximization in Hypergraphs using Multi-Objective Evolutionary Algorithms}
\titlerunning{Evolutionary Influence Maximization in Hypergraphs}

%

\author{
Stefano Genetti\thanks{Equal contribution.}\orcidID{0009-0004-2417-0319}
\and
Eros Ribaga\printfnsymbol{1}\orcidID{0009-0005-6345-6528}
\and\\
Elia Cunegatti\orcidID{0000-0002-1048-0373}
\and
Quintino F. Lotito\orcidID{0000-0001-7084-8339}
\and
Giovanni Iacca\orcidID{0000-0001-9723-1830}
}
\authorrunning{S. Genetti et al.}
%
\institute{University of Trento, Italy \\
\email{\{stefano.genetti,eros.ribaga\}@studenti.unitn.it}
\email{\{elia.cunegatti,quintino.lotito,giovanni.iacca\}@unitn.it}} 
\maketitle 
\begin{abstract}
The Influence Maximization (IM) problem is a well-known NP-hard combinatorial problem over graphs whose goal is to find the set of nodes in a network that spreads influence at most. Among the various methods for solving the IM problem, evolutionary algorithms (EAs) have been shown to be particularly effective. While the literature on the topic is particularly ample, only a few attempts have been made at solving the IM problem over higher-order networks, namely extensions of standard graphs that can capture interactions that involve more than two nodes. Hypergraphs are a valuable tool for modeling complex interaction networks in various domains; however, they require rethinking of several graph-based problems, including IM. In this work, we propose a multi-objective EA for the IM problem over hypergraphs that leverages smart initialization and hypergraph-aware mutation. While the existing methods rely on greedy or heuristic methods, to our best knowledge this is the first attempt at applying EAs to this problem. Our results over nine real-world datasets and three propagation models, compared with five baseline algorithms, reveal that our method achieves in most cases state-of-the-art results in terms of hypervolume and solution diversity.

\end{abstract}


\section{Introduction}
\label{sec:introduction}

Networks provide a valuable framework to model and analyze systems of interacting unities. Networks are typically represented as a graph, namely, a collection of nodes (the units of the system) connected via edges (the interactions between those units). Given their flexibility, networks have found applications in several domains, from the study of human behavior and cellular interactions to the assessment of the resilience and efficiency of technological systems~\cite{boccaletti2006complex}.   
However, conventional graph models may fail at capturing the full complexity and heterogeneity characterizing real-world networks. In fact, while empirical interactions can be described by many complex features (e.g., direction, weight, temporality, etc.), standard graphs usually associate only one feature with each edge. Moreover, graphs can only encode pairwise interactions, oversimplifying systems characterized by higher-order interactions, i.e., group interactions among three or more units~\cite{battiston2020networks, battiston2021physics}. Examples of such systems are scientific collaborations~\cite{patania2017shape}, people's face-to-face encounters~\cite{cencetti2021temporal}, and the brain~\cite{petri2014homological}. In order to model such higher-order interactions, hypergraphs~\cite{berge1973graphs}, rather than standard graphs, are needed. Hypergraphs are a generalization of graphs in which interactions are encoded into sets of arbitrary size, i.e., the hyperedges.

At the interplay between network structure and dynamics, a popular problem over graphs is the so-called Influence Maximization (IM). In this problem, the aim is to select a set of nodes from which the influence can be spread at most over the network~\cite{kempe2003maximizing}. Solving this problem exactly has been proven to be NP-hard. IM has been mainly studied in single-objective formulation, i.e., given a predefined number of nodes to be picked as starting seeds, the only objective is to maximize the spread (i.e., the number of influenced nodes). IM has been extensively studied in the context of standard graphs, yet, its application to higher-order networks is still limited. In hypergraphs, influence can spread through groups, impacting multiple units simultaneously and leading to non-linear behaviors. Hence, dynamical processes on hypergraphs are a more accurate model for many complex real-world dynamics, such as social influence in groups of friends, that are oversimplified by graph representations~\cite{battiston2020networks,antelmi2021social}. However, the expressive power of hypergraphs comes at the cost of having to generalize traditional graph problems and algorithms to the higher-order case. In this direction, extending and solving the IM problem in hypergraphs allows the analysis of data that inherently represent a hypergraph (e.g., scientific collaborations), for which propagation models are directly defined as higher-order, and that cannot be run on top of standard graphs. Moreover, IM on hypergraphs also allows for better modeling systems previously studied with graph approximation of higher-order dynamics, aiming for better identification of influential nodes.

In this work, we propose a Higher-Order Network Multi-Objective Evolutionary Algorithm (in short, \method), the first algorithm that employs Evolutionary Computation to solve the IM problem over higher-order networks. We also increase the problem complexity by designing a bi-objective formulation where the influence spread (to maximize) and the number of nodes in the starting seed set (to minimize) are jointly optimized. 
We design our method by adapting, for higher-order networks, a state-of-the-art Evolutionary Algorithm (EA) for IM~\cite{improving-multi-objective,cunegatti2024many}, also including IM-specific techniques such as smart initialization~\cite{Konotopska_2021} and graph-aware mutations~\cite{cunegatti2024many} to further boost the evolutionary process.
We compare \method w.r.t. the most recent baselines for IM on higher-order networks over three different propagation models, showing how our proposed method always shows comparable or better performance both in terms of hypervolume and solution diversity. 
In summary:
\begin{itemize}[label=$\bullet$,noitemsep,topsep=0pt,parsep=0pt,partopsep=0pt]
    \item We propose \method, a multi-objective EA designed to solve the IM problem over higher-order networks;
    \item We adapt smart initialization and hypergraph-aware mutations usually designed for standard graphs to be hypergraph-dependent;
    \item We test our approach w.r.t. to two standard (i.e., non-hypergraph-specific) baseline methods as well as three recent IM algorithms specifically designed for higher-order networks, over three propagation models;
    \item We show how our approach not only provides, in general, higher hypervolumes, but also finds sets of non-dominated solutions that are inherently more diverse than those found by the compared methods.
\end{itemize}



\section{Background}
\label{sec:background}
Recently, there has been a growing interest in characterizing hypergraphs, from micro-scale patterns~\cite{lee2020hypergraph, lotito2022higher, lotito2023exact}, to core-periphery organization~\cite{tudisco2023core}, community structure~\cite{contisciani2022inference, ruggeri2023community,lotito2024hyperlink}, backboning~\cite{musciotto2021detecting}, and centrality measures~\cite{benson2019three}. This is mainly motivated by the fact that hypergraphs can encode, without losing information, systems that display higher-order interactions, thus providing new insights into those systems' behavior. Formally, a \emph{hypergraph} is an ordered pair $H(V,E)$, where $V=\{v_1,\hdots,v_n\}$ is the set of nodes and $E=\{e_1,\hdots,e_m\}$ is the set of \emph{hyperedges}. Each hyperedge $e \in E$ is a set of nodes with a cardinality of at least $2$, i.e., $e \subseteq V$ and $|e| \geq 2$. For any given node $v\in V$, the set $E(v) \in E$ refers to the collection of hyperedges containing $v$. Additionally, a node $u$ belongs to the set of neighbors $\mathcal{N}(v)$ of a node $v$ if there exists at least one $e \in E$ such that $e$ contains both $v$ and $u$ ($E(v) \cap E(u) \neq \emptyset$). The \emph{degree} $d$ of a node $v$ corresponds to the cardinality of its set of neighbors, expressed as $d(v)=|\mathcal{N}(v)|$. Whereas, the \emph{hyperdegree} $d^H$ of a node $v$ is the number of hyperedges to which $v$ belongs.

\subsection{Propagation models}
\label{subsec:propagation_models}
Higher-order interactions in complex systems can significantly alter propagation dynamics~\cite{battiston2021physics, battiston2020networks}. Therefore, understanding how higher-order interactions affect different dynamical processes (e.g., contagion~\cite{iacopini2019simplicial, chowdhary2021simplicial} or synchronization~\cite{gambuzza2021stability}) previously studied in the traditional graph setting is attracting interest. In this work, our focus is on influence propagation. 

In standard graphs, influence propagation is typically modeled as an iterative process in which, given a graph $(V,E)$, at each iteration (timestep $t$), each node in $V$ can be either \emph{active} (i.e., it has been influenced) or \emph{inactive}. The influence spread starts at $t_0$ from a set of \emph{seed nodes} $S \subseteq V$, which are all active, while all the other nodes in the graph are not. Then, at each timestep, each of the active nodes can influence one or more of its neighbors, according to different logics (e.g., based on a certain probability) that depend on the given \emph{propagation model}. It is typically assumed that, once a node becomes active, it cannot become inactive anymore. Hence, the set of active nodes in $V$ increases monotonously over the timesteps, until the spreading process ends.

This general propagation process applies also to the case of hypergraphs of the form $H(V,E)$: what changes, is only the propagation model under which such process occurs. However, while propagation models devised for standard graphs are well-established, the literature on propagation models in higher-order networks is still limited. Developing propagation models tailored for hypergraphs that account for the complexity of hyperedges is nevertheless crucial for understanding and predicting influence propagation in diverse real-world scenarios that would be oversimplified with a traditional network representation and lower-order dynamics~\cite{battiston2020networks, battiston2021physics}.

In our experiments, we consider three different propagation models, namely the Weighted Cascade (WC)~\cite{kempe2003maximizing}, a model commonly adopted in the case of standard graphs and generalized to the higher-order domain in~\cite{zhu2018social}, as well as two recently-introduced hypergraph-specific models, referred to as Susceptible-Infected Contact Process (SICP)~\cite{xie2022influence} and Linear Threshold (LT)~\cite{zhang2024influence}.\\
%
\Circled{1} \noindent \textbf{Weighted Cascade (WC).} At each timestep $t \geq 1$, each node $n$ active at time $t-1$ may activate some of its inactive neighbors $m$ with non-uniform probability inversely proportional to the number of neighbors of $m$, i.e., the probability of $a \rightarrow b$ is given by $\frac{1}{d(b)}$.\\
%
\Circled{2} \noindent \textbf{Susceptible-Infected Contact Process (SICP).} At each timestep $t$, for each active node $n$, we consider all the hyperedges $E_i=\{e_{i1}, e_{i2}, \hdots e_{iq}\}$ in which node $n$ participates. At this point, a hyperedge $e$ is sampled from $E_i$ uniformly at random. Then, each of the inactive nodes in $e$ is influenced by node $n$ with probability $p$, i.e., the probability of $a \rightarrow b$ is given by $p$.\\
%
\Circled{3} \noindent \textbf{Linear Threshold (LT).} 
Let us consider a hyperedge $e$ 
with $A$ be the set of nodes in active state. 
In this scenario, a hyperedge $e$ becomes activated if the fraction of activated nodes $\frac{|A|}{|e|}$ is greater than or equal to the configured threshold value $\theta \in (0,1)$. Upon activation of $e$, all nodes $v$ belonging to $e$ also transition to an active state in the subsequent step of the propagation process. For instance, let $e_i=\{v_{i1}, v_{i2}, v_{i3}, v_{i4}\}$ denote a hyperedge comprising four nodes. Suppose that, at time $t$, the set $A_t$ includes $v_{i1}, v_{i3}$. Then, at time $t+1$, if the fraction of active nodes in $e_i$ exceeds $\theta$, $e_i$ will be activated. Consequently, all currently inactive vertices of $e_i$, namely $v_{i2}$ and $v_{i4}$, will also become active.
\smallskip

\begin{figure}[t!]
    \centering
    \includegraphics[width=.8\textwidth]{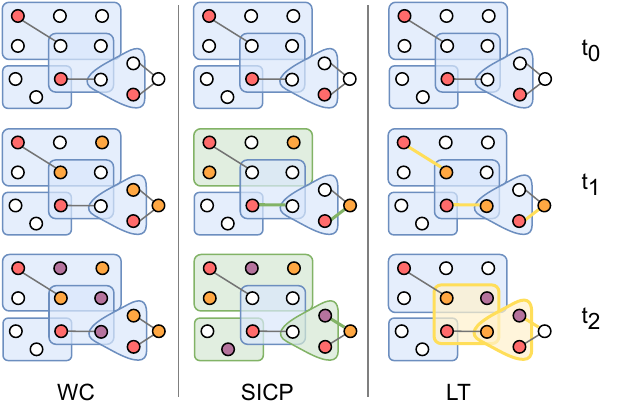}
    \caption{Graphical representation of the WC, SICP, and LT models. Each colored polygon indicates a hyperedge. The first row represents the hypergraph at time $t_0$, where only the nodes of the seed set (\textcolor[HTML]{FF6B6B}{Red}) are activated. The second row shows the nodes that are activated at timestep $t_1$ (\textcolor{BurntOrange}{Orange}), while the bottom row depicts the nodes that are activated at timestep $t_2$ (\textcolor[HTML]{B5739D}{Purple}). The second column displays the hyperedges that SICP randomly selects to spread the influence (\textcolor[HTML]{82B366}{Green}). The third column highlights the activation of hyperedges based on the LT propagation model, assuming a threshold value of $0.5$ (\textcolor[HTML]{FFDE59}{Yellow}).}
    \label{fig:propagationModels}
    \vspace{-0.5cm}
\end{figure}

We illustrate the considered propagation models in~\Cref{fig:propagationModels}. In all these models, the propagation of influence terminates either upon reaching convergence, indicated by no further nodes being activated in the last timestep, or upon reaching a specified maximum number of timesteps $\tau$ (maximum number of hops). Due to the stochasticity that characterizes the WC and SICP propagation models, the influence propagation is computed through multiple Monte Carlo simulations. On the other hand, the LT model has been designed to provide a deterministic execution~\cite{zhang2024influence}, without requiring multiple simulations.

\subsection{Influence Maximization problem}
\label{sec:problem}
Given a seed set $S$, its influence, denoted as $\sigma(S)$, is the (expected, in the case of stochastic propagation models) size of the set of active nodes at the end of the influence propagation process. Introduced in~\cite{domingos2001mining} and further formalized as a combinatorial optimization problem in~\cite{kempe2003maximizing}, the IM problem aims to identify the seed set of nodes $S$ in the network that maximizes the number of influenced nodes, i.e., $S = argmax_S\{\sigma(S)\}$. In the traditional formulation, the number of nodes $|S|$ in the seed set is predefined. As detailed in \Cref{sec:method}, in this work we highlight the importance of including $|S|$ as an additional objective of the optimization problem, to direct the search towards valuable trade-offs between \emph{effort} (number of seed nodes) and \emph{effect} (final influence over the whole network). Of note, such bi-objective formulation also leads to higher solution diversity among different seed set sizes, as we will demonstrate in the Results section.

\section{Related work}
\label{sec:rw}
The existing literature on IM is mainly focused on standard graphs~\cite{rahimkhani2015fast, nguyen2016stop, goyal2011celf++, ling2023deep}. Over the years, researchers have explored various algorithms to address this problem~\cite{banerjee2020survey, li2018influence, peng2018influence, li2023survey}. Among the proposed solutions, EAs and other forms of metaheuristics~\cite{lotf2022improved, cui2018ddse, tang2021improved, zhang2017maximizing, kromer2018guided} have demonstrated remarkable effectiveness in tackling this kind of combinatorial optimization problem~\cite{aghaee2021survey}. For instance,~\cite{gong2016influence} proposed the local influence estimation (LIE) function, which considers the influence within the 2-hop neighborhood of seed nodes, and optimized it using the Discrete Particle Swarm Optimization (DPSO) algorithm. In~\cite{jiang2011simulated}, the authors introduced the expected diffusion value (EDV) evaluation function and utilized Simulated Annealing (SA) to identify the most influential nodes. Moreover, recent research has showcased the efficacy of single-objective~\cite{bucur2016influence} and multi-objective EAs~\cite{moea-for-im-in-social-networks, improving-multi-objective} in outperforming alternative approaches both in terms of quality and execution times~\cite{cunegatti2022large}. Notably, the method outlined in~\cite{moea-for-im-in-social-networks,cunegatti2022large} deviates from the traditional formulation of the problem, where the cardinality of the seed set is predefined. Setting the seed set size cardinality \emph{a priori} indeed inherently restricts the solution space explored by the optimization process. In line with this observation, the multi-objective formulation in~\cite{moea-for-im-in-social-networks,cunegatti2024many} aims to simultaneously maximize the collective influence while minimizing the size of the seed set, possibly along with other objectives such as the propagation time, the influence fairness, or the cost of propagation, while improving solution diversity.


While the IM problem has been extensively studied in the context of lower-order interactions, IM in the higher-order case remains relatively unexplored. In principle, the IM problem in higher-order networks can be solved by applying existing IM algorithms designed for standard graphs on the hypergraph's clique-expanded graph, i.e., a graph representation in which edges connect vertices that are part of the same hyperedge in the original hypergraph. Although executing an IM algorithm on clique-expanded graphs is a viable strategy for identifying influential sources, it is important to recognize that this process inevitably sacrifices several crucial topological features of the original hypergraph data structure~\cite{wang2024graphs}. In more depth, in a higher-order network, multiple interactions could potentially be shared by two neighboring nodes. On the other hand, in its low-order graph counterpart, each pair of nodes can only have one pairwise interaction. This distinction significantly impacts the spread of the influence across the network~\cite{battiston2020networks}. Moreover, the insofar proposed propagation models for hypergraphs are specifically tailored for higher-order networks, and may not perform well when applied to networks with only dyadic interactions.

Given that IM is NP-hard also on higher-order networks~\cite{zhu2018social}, existing works~\cite{antelmi2021social, zhang2024influence} rely on greedy~\cite{zheng2019non,wang2024hedv} or heuristic strategies~\cite{xie2023efficient} to explore the search space within reasonable computational time. Overall, these methods evaluate the suitability of each node as a source of influence spread by assigning it a score, and incrementally add nodes with the highest marginal benefit to the seed set. These techniques have effectively addressed the IM problem across various real-world higher-order network datasets. However, as we will show in our experiments, they are characterized by limited exploration capabilities and resulting solution diversity. In contrast to standard graphs, no prior work has yet explored the resolution of the IM problem in hypergraphs using EAs, which instead can provide better exploration and diversity.


\section{Methodology}
\label{sec:method}

In agreement with the methodology originally suggested in~\cite{moea-for-im-in-social-networks}, our formulation of the IM problem does not enforce any \emph{a priori} constraint on the cardinality of the seed set. Instead, the multi-objective formulation in this study aims to maximize the collective influence while minimizing the size of the seed set $|S|$. Following this methodology, given an input hypergraph $H(V,E)$, the genotype of an individual $x$ generated throughout the evolutionary process encodes a set of nodes $S \subseteq V$ of variable size in $\{1, 2, \dots, k\}$, representing the seeds of influence in the network. Each node is indicated by its id, i.e., an integer in $\{0,1, \dots,|V|-1\}$. The fitness of a candidate solution $x$ is a tuple containing: (i) the influence $\sigma(x)$ of the seed set, calculated as described in \Cref{sec:problem}, to be maximized $\uparrow$; (ii) the size $k$ of the seed set $S$, to be minimized $\downarrow$. Both values are normalized w.r.t. the network size, to allow comparisons between networks of different sizes.

The multi-objective EA of choice in this work is NSGA-II~\cite{deb2002fast}, which has been proven to be successful on the IM problem in standard graphs, outperforming in most cases the alternative heuristics~\cite{cunegatti2022large}. Moreover, this method can be easily extended to incorporate additional objective functions into the optimization process, as shown in~\cite{cunegatti2024many}. 
We also use the smart initialization strategies proposed in~\cite{Konotopska_2021,chawla2023neighbor,cunegatti2024many}, 
aiming at accelerating algorithm execution and guiding population convergence towards prominent regions of the solution space. 
In line with the strategy adopted in~\cite{moea-for-im-in-social-networks, improving-multi-objective, cunegatti2022large}, parent solutions are selected with fixed-size tournament and elitism. The offspring solutions are generated by standard one-point crossover, while for mutation we took inspiration from the graph-based mutation presented in~\cite{Konotopska_2021,cunegatti2024many}.
For individual replacement, we rely on the standard NSGA-II replacement mechanism consisting of non-dominated sorting followed by crowding distance preference.

\noindent \textbf{Smart Initialization.}
Generating an initial population situated within prominent regions of the solution space is a practice commonly adopted in Evolutionary Computation~\cite{6557692, 6557763, 5953633} in order to facilitate convergence towards profitable fitness landscape regions. 
This approach has been proven to be effective also on the IM problem over standard graphs~\cite{Konotopska_2021,cunegatti2024many}. 
Hence, we decided to adopt the same approach also on the hypergraphs handled in this paper. 
The rationale behind this approach is rooted in the observation that nodes characterized by high centrality are likely to be effective sources of influence spread. 

Our smart initialization over higher-order networks works as follows: initially, given a hypergraph $H(V,E)$, we sort the nodes in the set $V$ based on their degree. Subsequently, we select the top $\lambda$ percentage of nodes with the highest degree, hence obtaining a filtered set of nodes $\Bar{V}$ (with $\lambda$ being a hyperparameter).
Half of the initial population consists of a set of nodes sampled from the filtered set $\Bar{V}$, with probabilities proportional to their degrees. In order to favor diversity in the population, the other half comprises seed sets of nodes chosen uniformly at random from the entire node set $V$. To further promote diversity, each individual's genotype in the initial population is initialized with a randomly selected number of nodes, ranging from $k_{\text{min}}$ to $k_{\text{max}}$ (both are hyperparameters). 

\noindent \textbf{Hypergraph-aware mutation.}
While the employment of random mutation and one-point crossover leads to remarkable performance on the IM problem~\cite{improving-multi-objective, cunegatti2022large}, introducing hypergraph-aware mutation operators can guide the evolutionary process towards even better results~\cite{Konotopska_2021, cunegatti2024many}. Hence, we rely on a combination of stochastic and hypergraph-aware mutation operators, in order to strike a balance between exploration and exploitation. Each individual is mutated according to one of the two mutation operators, selected uniformly at random.\\
%
\Circled{1} \noindent \textbf{Stochastic mutation.} 
Given an individual $x$ (i.e., a seed set), with a genotype consisting of $l$ genes, this mutation performs either: (1) \emph{node replacement}, which generates a new individual $x'$ by randomly replacing one of the genes of $x$ with a node $n \notin x$; (2) \emph{node insertion}, which generates a new individual $x'$ with $l+1$ genes by adding to $x$ a new node $n \notin x$; or (3) \emph{node removal}, which generates a new individual $x'$ with $l-1$ genes, by randomly removing from $x$ a node $n \in x$. The three strategies are chosen at random with uniform probability.\\
%
\Circled{2} \noindent \textbf{Hypergraph-aware mutation.} This mutation leverages the intrinsic characteristics of nodes. In this case, we only consider \emph{node replacement}. Given again an individual $x$ of size $l$, first, we select the gene to be replaced with a probability inversely proportional to its corresponding node degree. Then, we choose (with equal probability) the new node $n \notin x$ either from the entire collection of nodes $V$, or from the neighbors of the gene selected for replacement. In both cases, the new node is chosen with probability proportional to its degree.


\section{Experimental setup}
\label{sec:experiment}

In this section, we provide an overview of the experimental design employed to evaluate the efficacy of \method, as well as the tested baselines.

\noindent
\textbf{Computational setup.} We performed our experiments on two Ubuntu 20.04 workstations, respectively with a 28-core Intel i9-7940X CPU @ 3.10GHz and 64GB RAM, and a 36-core Intel i9-10980XE CPU @ 3.00GHz and 128GB RAM. The total execution time of our experiments was in the order of (approximately) 400 CPU core hours. 
The code implementing our methods is completely written in Python and is made publicly available\footnote{\url{https://github.com/DIOL-UniTN/hn-moea-im.git}}. Furthermore, it has been integrated into the Hypergraphx library~\cite{lotito2023hypergraphx}.

\noindent
\textbf{Datasets.}
\label{subsec:datasets}
In order to evaluate the effectiveness of our proposed method, and to allow for a direct comparison with other methods for IM on hypergraphs, we performed an experimental analysis on nine publicly available real-world higher-order networks\footnote{\url{https://www.cs.cornell.edu/~arb/data/}} used in related works. The selected datasets represent empirical hypergraphs from three heterogeneous domain categories spanning social networks, online reviews, and email communication. These datasets cover a wide range of different topological properties, i.e., number of nodes, number of hyperedges, and density. Each dataset has been properly pre-processed to remove duplicated hyperedges, duplicated nodes within the same higher-order interaction, and relations populated by less than two entities. Summary statistics of the datasets, after pre-processing, are available in \Cref{tab:hypergraph_datasets}.

\begin{table}[ht!]
\caption{Datasets tested in our experimental setup, divided by category: social (Algebra, Geometry, MAG-10), online reviews (Restaurant, Music, Bars), and email communication (Email-eu, Email-enron, Email-w3c).}
\label{tab:hypergraph_datasets}
\resizebox{\textwidth}{!}{
\begin{tabular}{l r r rrr rrr c}
\toprule
\multirow{2}[2]{*}{\textbf{Dataset}} & \multirow{2}[2]{*}{\textbf{Nodes}} & \multirow{2}[2]{*}{\textbf{Hyperedges}} & \multicolumn{3}{c}{\textbf{Hyperdegree}} & \multicolumn{3}{c}{\textbf{Degree}} & \multirow{2}[2]{*}{\textbf{Source}} \\ \arrayrulecolor{gray} \cmidrule(lr){4-6} \cmidrule(lr){7-9} \arrayrulecolor{black}
& & & \textbf{Avg.} & \textbf{Std.}& \textbf{Max.} & \textbf{Avg.} & \textbf{Std.} & \textbf{Max.} & \\
\midrule
Algebra & 423 &980 & 17.52 & 29.93 & 328 & 78.89 & 68.38 & 303 &~\cite{amburg2020fair} \\
Geometry & 580 &888 & 19.90 & 32.69 & 227 & 164.79 & 121.62 & 474 &~\cite{amburg2020fair}\\
MAG-10& 80198& 51889& 2.25& 4.56& 187& 5.91& 9.19& 335&~\cite{Amburg-2020-categorical, Sinha-2015-MAG}\\
\midrule
Restaurant & 565 &594 & 8.11 & 7.17 & 59 & 79.75 & 59.82 & 310 &~\cite{amburg2020fair}\\
Music & 1106 &686 & 9.47 & 10.72 & 127 & 167.87 & 107.92 & 865 &~\cite{ni2019justifying}\\
Bars & 1234 &1188 & 9.60 & 7.36 & 146 & 174.30 & 145.02 & 818 &~\cite{amburg2020fair}\\
\midrule
Email-eu& 1000& 78919& 259.12& 340.84& 2386& 280.44& 217.53& 755&~\cite{Benson-2018-simplicial,Yin-2017-local,Leskovec-2007-evolution} \\
Email-enron& 4423&5734& 6.80& 32.05& 1139& 25.34& 43.96& 934&~\cite{Benson-2018-simplicial} \\
Email-w3c& 14317& 19821& 3.07& 23.90& 958& 4.06& 23.98& 959&~\cite{amburg2021planted, Craswell-2005-TREC}\\
\bottomrule
\end{tabular}
}
\vspace{-0.5cm}
\end{table}

\noindent \textbf{Baselines.}
\label{subsec:baselines}
We compare our proposed \method with five other algorithms for IM in higher-order networks. Of these, two are non-hypergraph-specific (\textsc{random} and \textsc{high-degree}), while the other three (\textsc{hdd}, \textsc{hci-1}, and \textsc{hci-2}) are specific for hypergraphs. Furthermore, four of the compared methods are deterministic (\textsc{high-degree}, \textsc{hdd}, \textsc{hci-1}, and \textsc{hci-2}) while \textsc{random} is stochastic. In the following, we consider a hypergraph $H(V,E)$ and a maximum seed set size $k_{\text{max}}$.

The \textsc{random} algorithm simply generates $k_{\text{max}}$ seed sets, with sizes ranging from $k_{\text{min}}$ to $k_{\text{max}}$, by iteratively adding a node randomly sampled from $V$.

In the \textsc{high-degree} approach~\cite{xie2022influence}, nodes in $V$ are sorted according to their degree. The output Pareto Front comprises candidate seed sets $S_1, \hdots, S_{k_{\text{max}}}$ of increasing sizes from $k_{\text{min}}$ to $k_{\text{max}}$. For each set size $i$, the top $k_i$ nodes with the highest degree are selected ($S_i = argmax_{S' \subseteq V, |S'|=k_i}\sum_{v \in S'}d(v)$).

Along with these two non-hypergraph-specific baselines, we include in our experiments three of the most recent IM algorithms for hypergraphs proposed in the literature. Specifically, we consider \textsc{hdd}, proposed in~\cite{xie2022influence}, as well as the \textsc{hci-1} and \textsc{hci-2} algorithms introduced in~\cite{zhang2024influence}. Of note, all these methods have been proposed to be single-objective. Hence, to compare them with \method we executed them for every value of seed set size $k$ within the interval $[k_{\text{min}}, k_{\text{max}}]$.
\smallskip

\noindent \textbf{Hyperparameter setting.}
\label{subsec:hyperparameter_setting}
To strike a balance between computational efficiency and accurately capturing the influence spread dynamics, we set the maximum number of hops within which influence is propagated to $\tau=5$ as in~\cite{gong2023influence,cunegatti2024many}, which reflects a fair compromise between the commonly adopted 2-hop approximation and an unbounded spread process (i.e., $\tau=\infty$). 

To limit the hyperparameter dependency for the SICP model, rather than relying on a constant value for the probability $p$, we sample $p$ at each timestep uniformly at random within $[0.005, 0.02]$ (values commonly used in~\cite{xie2022influence}). 
Regarding the LT propagation model, the spread of influence is extremely sensitive to the threshold $\theta$ and it is not feasible to identify a value for this parameter that is suitable for every dataset. Therefore, we opted for a different threshold for each network. Specifically, we adopted the values utilized in~\cite{zhang2024influence}: $\theta=0.8$ for {Algebra} and {Geometry}, $\theta=0.5$ for {MAG-10}, $\theta=0.6$ for {Music} and {Bars}. For the remaining datasets, we employed the approach outlined in~\cite{zhang2024influence}, wherein the parameter $\theta$ is tailored to the specific characteristics of each dataset. Following this strategy, through empirical investigation, we determined $\theta=0.7$ for {Restaurant} and {Email-enron}, $\theta=0.6$ for {Email-w3c}, and $\theta=0.8$ for {Email-eu}. 

Concerning the EA parameters, as proposed in~\cite{improving-multi-objective,cunegatti2024many} we set the minimum and the maximum seed set size of an individual to $k_{\text{min}}=1$ and $k_{\text{max}}=100$ respectively. The parameter $\lambda$ used in our smart initialization strategy has been set to $\lambda=30\%$. For all the experiments, the evolutionary hyperparameters have been kept fixed, setting the population size to $100$, number of offspring to $100$, number of elites to $2$, tournament size to $5$, and generations to $100$ (as in~\cite{cunegatti2024many}). When evaluating the fitness of the \method solutions w.r.t. WC and SICP, we conduct $100$ Monte Carlo simulations of the propagation model, while the LT model~\cite{zhang2024influence} being deterministic does not require multiple evaluations. 
To deal with the inherent stochasticity of \method and the \textsc{random} baseline, the results presented below for these two algorithms have been aggregated from $5$ independent runs, providing a more robust and reliable assessment of outcomes. Conversely, due to their deterministic nature, the results for the other baselines (\textsc{high-degree}, \textsc{hdd}, \textsc{hci-1}, \textsc{hci-2}) are based on a single execution.

The hyperparameters of the \textsc{hci-1} and \textsc{hci-2} algorithms~\cite{zhang2024influence} have been carefully fine-tuned. Indeed, due to the specific characteristics of \textsc{hci-1} and \textsc{hci-2}, the selection of certain parameters significantly impacts their ability to identify seed sets for different values of $k$ within the range of interest, namely $[1,100]$. In more depth, we adjusted the hyperedge threshold parameter in their source code to $0.85$, which ensures fair results across all datasets analyzed in our study.


\section{Results}
\label{sec:results}

We analyze the performance of \method w.r.t. the compared algorithms for IM both in terms of hypervolume and solution diversity. 

\begin{figure}[ht!]
    \centering
    \includegraphics[width=.85\textwidth]{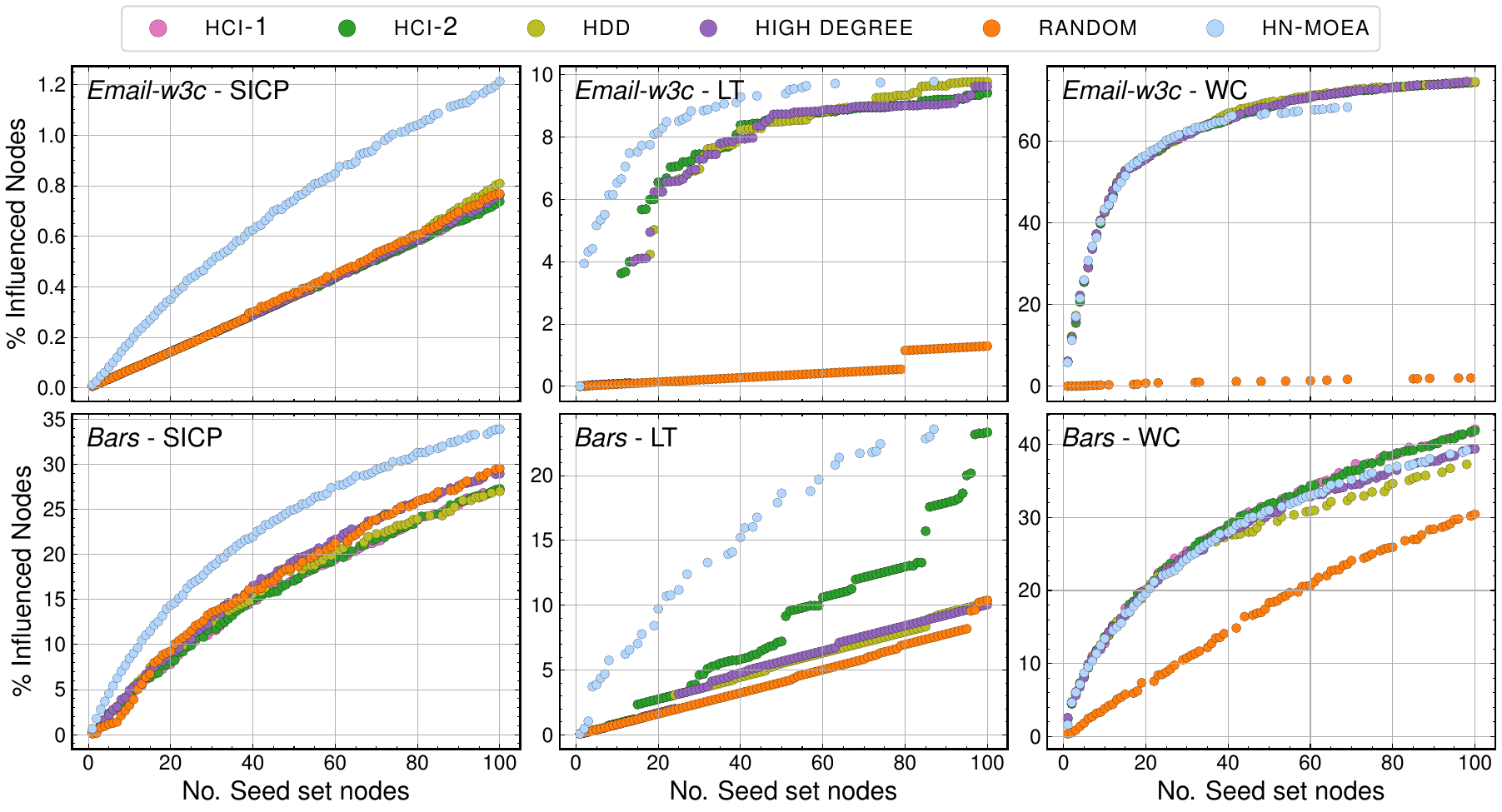}
    \caption{Results obtained by the compared IM algorithms on the Email-w3c and Bars datasets, using WC, SICP, and LT as influence propagation models.}
    \label{fig:paretoFronts}
    \vspace{-0.7cm}
\end{figure}

\noindent \textbf{Performance in terms of hypervolume}
We compare our proposed approach and the baselines both qualitatively and quantitatively. \Cref{fig:paretoFronts} displays a qualitative representation of the performance attained by the evaluated IM algorithms on two selected datasets, considering all three influence propagation models examined in this study. 
Remarkably, in most cases the Pareto Fronts generated by \method demonstrate superior performance compared to the solutions obtained by other algorithms, achieving more favorable trade-offs between seed set size and percentage of influenced nodes. 
However, in consonance with the observations from~\cite{moea-for-im-in-social-networks}, for some datasets, NSGA-II struggles in populating the Pareto Front for solutions with node counts approaching the upper bound of the seed set size. This might be due to the fact that, while for larger $k$ there exist less possible combinations, finding them becomes harder.
%


\Cref{tab:hypervolumeComparison} provides a quantitative comparison of the algorithmic solutions by computing the hypervolume~\cite{shang2020survey} under the curve of the Pareto Front found by the various methods on each dataset and propagation model. 
Here, it can be seen that \method excels particularly within propagation models tailored for the higher-order domain (SICP and LT), often outperforming competitors by a significant margin. On the other hand, in the case of a propagation model initially intended for standard graphs (WC), our algorithm's performance does not exhibit a notable enhancement compared to the evaluated baselines. Nevertheless, in these instances, the quality of the solutions proposed by our method aligns closely with that of other algorithms. These observations hold considerable importance in practical applications, as real-world scenarios often encompass diverse influence propagation patterns. Consequently, it is essential to develop algorithmic solutions capable of producing valuable outcomes across a wide range of propagation models.
Furthermore, \method demonstrates remarkable versatility not only with respect to the propagation model but also across the diverse datasets analyzed in this work, being able to effectively converge towards profitable solutions regardless of the peculiarities inherent to the different network domains. However, it is worth noting that its performance on the MAG-10 dataset is slightly less satisfactory. This can be attributed to the large number of nodes and hyperedges present in this dataset, resulting in a significantly expanded solution space. Allowing the algorithm to evolve over more generations and with a larger population size could potentially yield better outcomes.

\definecolor{verylightpastelskyblue}{HTML}{CCE6FF}

\begin{table}[ht!]
    \scriptsize
    \centering
    \caption{Hypervolumes achieved by the compared algorithms. Results for \textsc{random} and \method are cross $5$ independent runs (mean $\pm$ std. dev.). The boldface indicates the highest hypervolume per dataset and propagation model. We highlight the cases where our method achieves the highest hypervolume.}
    \label{tab:hypervolumeComparison}
    \resizebox{.86\textwidth}{!}{
    \begin{tabular}{ l l c c c}
    \toprule
    \multirow{2}[2]{*}{\textbf{Dataset}} & \multirow{2}[2]{*}{\textbf{Algorithm}} & \multicolumn{2}{c}{\textbf{Higher-Order Propagation}} & \textbf{Standard Propagation} \\
    \cmidrule(l{5pt}r{5pt}){3-4} 
    \cmidrule(l{5pt}r{5pt}){5-5}
    & & \textbf{SICP} & \textbf{LT} & \textbf{WC} \\
    \midrule
    \multirow{6}[1]{*}{Algebra} & \textsc{high-degree} & $1.79\mathrm{e}{-01}$ & $4.67\mathrm{e}{-01}$ & $4.96\mathrm{e}{-01}$\\
    & \textsc{hdd} & $1.84\mathrm{e}{-01}$ & $1.65\mathrm{e}{-01}$ & $4.41\mathrm{e}{-01}$\\
    & \textsc{hci-1} & $1.67\mathrm{e}{-01}$ & $4.73\mathrm{e}{-01}$ & $4.69\mathrm{e}{-01}$\\
    & \textsc{hci-2} & $1.72\mathrm{e}{-01}$ & $3.35\mathrm{e}{-01}$ & $4.37\mathrm{e}{-01}$\\
    & \textsc{random} & $2.15\mathrm{e}{-01} \pm 5.85\mathrm{e}{-03}$ & $1.29\mathrm{e}{-01} \pm 1.58\mathrm{e}{-02}$ & $3.53\mathrm{e}{-01} \pm 1.10\mathrm{e}{-02}$\\
    & \method & \colorbox{verylightpastelskyblue}{$\mathbf{2.96\mathrm{e}{-01}\pm 7.86\mathrm{e}{-04}}$} & \colorbox{verylightpastelskyblue}{$\mathbf{5.28\mathrm{e}{-01} \pm 9.10\mathrm{e}{-03}}$} & \colorbox{verylightpastelskyblue}{$\mathbf{5.01\mathrm{e}{-01} \pm 1.12\mathrm{e}{-03}}$}\\
    \midrule
    \multirow{6}[1]{*}{Geometry} & \textsc{high-degree} & $2.39\mathrm{e}{-01}$ & $2.04\mathrm{e}{-01}$ & $\mathbf{4.34\mathrm{e}{-01}}$\\
    & \textsc{hdd} & $2.28\mathrm{e}{-01}$ & $1.10\mathrm{e}{-01}$ & $3.73\mathrm{e}{-01}$\\
    & \textsc{hci-1} & $2.22\mathrm{e}{-01}$ & $3.06\mathrm{e}{-01}$ & $4.24\mathrm{e}{-01}$\\
    & \textsc{hci-2} & $2.25\mathrm{e}{-01}$ & $2.50\mathrm{e}{-01}$ & $4.08\mathrm{e}{-01}$\\
    & \textsc{random} & $3.29\mathrm{e}{-01} \pm 7.09\mathrm{e}{-03}$ & $8.74\mathrm{e}{-02} \pm 7.31\mathrm{e}{-04}$ & $3.06\mathrm{e}{-01} \pm 6.04\mathrm{e}{-03}$\\
    & \method & \colorbox{verylightpastelskyblue}{$\mathbf{4.59\mathrm{e}{-01}\pm 1.86\mathrm{e}{-03}}$} & \colorbox{verylightpastelskyblue}{$\mathbf{3.06\mathrm{e}{-01} \pm 1.04\mathrm{e}{-02}}$} & $4.32\mathrm{e}{-01} \pm 8.67\mathrm{e}{-04}$\\
    \midrule
    \multirow{6}[1]{*}{MAG-10} & \textsc{high-degree} & $7.44\mathrm{e}{-04}$ & $2.63\mathrm{e}{-01}$ & $3.73\mathrm{e}{-02}$\\
    & \textsc{hdd} & $7.45\mathrm{e}{-04}$ & $2.65\mathrm{e}{-01}$ & $\mathbf{3.81\mathrm{e}{-02}}$\\
    & \textsc{hci-1} & $7.33\mathrm{e}{-04}$ & $\mathbf{2.76\mathrm{e}{-01}}$ & $3.72\mathrm{e}{-02}$\\
    & \textsc{hci-2} & $7.33\mathrm{e}{-04}$ & $\mathbf{2.76\mathrm{e}{-01}}$ & $3.71\mathrm{e}{-02}$\\
    & \textsc{random} & $7.42\mathrm{e}{-04} \pm 9.25\mathrm{e}{-06}$ & $2.18\mathrm{e}{-02} \pm 1.43\mathrm{e}{-02}$ & $2.31\mathrm{e}{-03} \pm 1.33\mathrm{e}{-04}$\\
    & \method & \colorbox{verylightpastelskyblue}{$\mathbf{1.30\mathrm{e}{-03}\pm 1.65\mathrm{e}{-05}}$} & $2.62\mathrm{e}{-01} \pm 2.73\mathrm{e}{-03}$ & $2.76\mathrm{e}{-02} \pm 8.97\mathrm{e}{-04}$\\
    \midrule
    \multirow{6}[1]{*}{Restaurant} & \textsc{high-degree} & $1.62\mathrm{e}{-01}$ & $1.38\mathrm{e}{-01}$ & $4.27\mathrm{e}{-01}$\\
    & \textsc{hdd} & $1.69\mathrm{e}{-01}$ & $8.95\mathrm{e}{-02}$ & $3.97\mathrm{e}{-01}$\\
    & \textsc{hci-1} & $1.51\mathrm{e}{-01}$ & $1.98\mathrm{e}{-01}$ & $\mathbf{4.33\mathrm{e}{-01}}$\\
    & \textsc{hci-2} & $1.51\mathrm{e}{-01}$ & $1.89\mathrm{e}{-01}$ & $4.32\mathrm{e}{-01}$\\
    & \textsc{random} & $1.72\mathrm{e}{-01} \pm 2.93\mathrm{e}{-03}$ & $8.98\mathrm{e}{-02} \pm 2.78\mathrm{e}{-04}$ & $3.06\mathrm{e}{-01} \pm 8.39\mathrm{e}{-03}$\\
    & \method & \colorbox{verylightpastelskyblue}{$\mathbf{2.18\mathrm{e}{-01}\pm 1.30\mathrm{e}{-03}}$} & \colorbox{verylightpastelskyblue}{$\mathbf{2.54\mathrm{e}{-01} \pm 1.35\mathrm{e}{-02}}$} & $4.31\mathrm{e}{-01} \pm 1.00\mathrm{e}{-03}$\\
    \midrule
    \multirow{6}[1]{*}{Music} & \textsc{high-degree} & $1.65\mathrm{e}{-01}$ & $1.80\mathrm{e}{-01}$ & $3.10\mathrm{e}{-01}$\\
    & \textsc{hdd} & $1.98\mathrm{e}{-01}$ & $6.98\mathrm{e}{-02}$ & $2.89\mathrm{e}{-01}$\\
    & \textsc{hci-1} & $1.38\mathrm{e}{-01}$ & $\mathbf{3.38\mathrm{e}{-01}}$ & $2.98\mathrm{e}{-01}$\\
    & \textsc{hci-2} & $1.38\mathrm{e}{-01}$ & $3.37\mathrm{e}{-01}$ & $2.99\mathrm{e}{-01}$\\
    & \textsc{random} & $2.10\mathrm{e}{-01} \pm 6.88\mathrm{e}{-03}$ & $5.92\mathrm{e}{-02} \pm 1.74\mathrm{e}{-02}$ & $1.95\mathrm{e}{-01} \pm 6.97\mathrm{e}{-03}$\\
    & \method & \colorbox{verylightpastelskyblue}{$\mathbf{2.94\mathrm{e}{-01}\pm 9.13\mathrm{e}{-04}}$} & $3.21\mathrm{e}{-01} \pm 2.16\mathrm{e}{-02}$ & \colorbox{verylightpastelskyblue}{$\mathbf{3.10\mathrm{e}{-01} \pm 5.28\mathrm{e}{-04}}$}\\
    \midrule
    \multirow{6}[1]{*}{Bars} & \textsc{high-degree} & $1.77\mathrm{e}{-01}$ & $5.40\mathrm{e}{-02}$ & $2.81\mathrm{e}{-01}$\\
    & \textsc{hdd} & $1.67\mathrm{e}{-01}$ & $5.25\mathrm{e}{-02}$ & $2.70\mathrm{e}{-01}$\\
    & \textsc{hci-1} & $1.62\mathrm{e}{-01}$ & $8.79\mathrm{e}{-02}$ & $\mathbf{2.94\mathrm{e}{-01}}$\\
    & \textsc{hci-2} & $1.62\mathrm{e}{-01}$ & $8.79\mathrm{e}{-02}$ & $2.93\mathrm{e}{-01}$\\
    & \textsc{random} & $1.72\mathrm{e}{-01} \pm 9.54\mathrm{e}{-03}$ & $4.52\mathrm{e}{-02} \pm 3.78\mathrm{e}{-03}$ & $1.73\mathrm{e}{-01} \pm 6.29\mathrm{e}{-03}$\\
    & \method & \colorbox{verylightpastelskyblue}{$\mathbf{2.28\mathrm{e}{-01}\pm 4.04\mathrm{e}{-04}}$} & \colorbox{verylightpastelskyblue}{$\mathbf{1.54\mathrm{e}{-01} \pm 1.44\mathrm{e}{-02}}$} & $2.82\mathrm{e}{-01} \pm 1.16\mathrm{e}{-03}$\\
    \midrule
    \multirow{6}[1]{*}{Email-eu} & \textsc{high-degree} & $6.03\mathrm{e}{-02}$ & $7.90\mathrm{e}{-01}$ & $\mathbf{3.48\mathrm{e}{-01}}$\\
    & \textsc{hdd} & $6.47\mathrm{e}{-02}$ & $7.91\mathrm{e}{-01}$ & $2.93\mathrm{e}{-01}$\\
    & \textsc{hci-1} & $5.87\mathrm{e}{-02}$ & $8.05\mathrm{e}{-01}$ & $2.91\mathrm{e}{-01}$\\
    & \textsc{hci-2} & $5.79\mathrm{e}{-02}$ & $8.22\mathrm{e}{-01}$ & $2.50\mathrm{e}{-01}$\\
    & \textsc{random} & $6.25\mathrm{e}{-02} \pm 5.57\mathrm{e}{-04}$ & $5.28\mathrm{e}{-02} \pm 4.10\mathrm{e}{-03}$ & $2.19\mathrm{e}{-01} \pm 8.82\mathrm{e}{-03}$\\
    & \method & \colorbox{verylightpastelskyblue}{$\mathbf{7.25\mathrm{e}{-02}\pm 5.21\mathrm{e}{-04}}$} & \colorbox{verylightpastelskyblue}{$\mathbf{8.36\mathrm{e}{-01} \pm 1.29\mathrm{e}{-02}}$} & $3.40\mathrm{e}{-01} \pm 1.21\mathrm{e}{-03}$\\
    \midrule
    \multirow{6}[1]{*}{Email-enron} & \textsc{high-degree} & $1.58\mathrm{e}{-02}$ & $2.41\mathrm{e}{-01}$ & $3.63\mathrm{e}{-01}$\\
    & \textsc{hdd} & $1.69\mathrm{e}{-02}$ & $2.45\mathrm{e}{-01}$ & $\mathbf{3.95\mathrm{e}{-01}}$\\
    & \textsc{hci-1} & $1.59\mathrm{e}{-02}$ & $\mathbf{2.60\mathrm{e}{-01}}$ & $3.85\mathrm{e}{-01}$\\
    & \textsc{hci-2} & $1.69\mathrm{e}{-02}$ & $2.41\mathrm{e}{-01}$ & $3.56\mathrm{e}{-01}$\\
    & \textsc{random} & $1.94\mathrm{e}{-02} \pm 5.38\mathrm{e}{-04}$ & $1.14\mathrm{e}{-02} \pm 0.00\mathrm{e}{+00}$ & $5.26\mathrm{e}{-02} \pm 5.49\mathrm{e}{-03}$\\
    & \method & \colorbox{verylightpastelskyblue}{$\mathbf{2.85\mathrm{e}{-02}\pm 2.15\mathrm{e}{-04}}$} & $2.29\mathrm{e}{-01} \pm 4.12\mathrm{e}{-03}$ & $3.91\mathrm{e}{-01} \pm 4.03\mathrm{e}{-03}$\\
    \midrule
    \multirow{6}[1]{*}{Email-w3c} & \textsc{high-degree} & $3.72\mathrm{e}{-03}$ & $7.13\mathrm{e}{-02}$ & $6.32\mathrm{e}{-01}$\\
    & \textsc{hdd} & $3.80\mathrm{e}{-03}$ & $7.21\mathrm{e}{-02}$ & $\mathbf{6.35\mathrm{e}{-01}}$\\
    & \textsc{hci-1} & $3.67\mathrm{e}{-03}$ & $7.32\mathrm{e}{-02}$ & $6.31\mathrm{e}{-01}$\\
    & \textsc{hci-2} & $3.67\mathrm{e}{-03}$ & $7.32\mathrm{e}{-02}$ & $6.31\mathrm{e}{-01}$\\
    & \textsc{random} & $3.76\mathrm{e}{-03} \pm 7.57\mathrm{e}{-05}$ & $3.53\mathrm{e}{-03} \pm 4.34\mathrm{e}{-19}$ & $1.14\mathrm{e}{-02} \pm 8.84\mathrm{e}{-04}$\\
    & \method & \colorbox{verylightpastelskyblue}{$\mathbf{7.03\mathrm{e}{-03}\pm 4.29\mathrm{e}{-05}}$} & \colorbox{verylightpastelskyblue}{$\mathbf{8.79\mathrm{e}{-02} \pm 1.94\mathrm{e}{-03}}$} & $6.04\mathrm{e}{-01} \pm 8.58\mathrm{e}{-03}$\\
    \bottomrule
    \end{tabular}
    }
    \vspace{-0.75cm}
\end{table}

\begin{figure}[ht!]
    \centering
    \includegraphics[width=.9\textwidth]{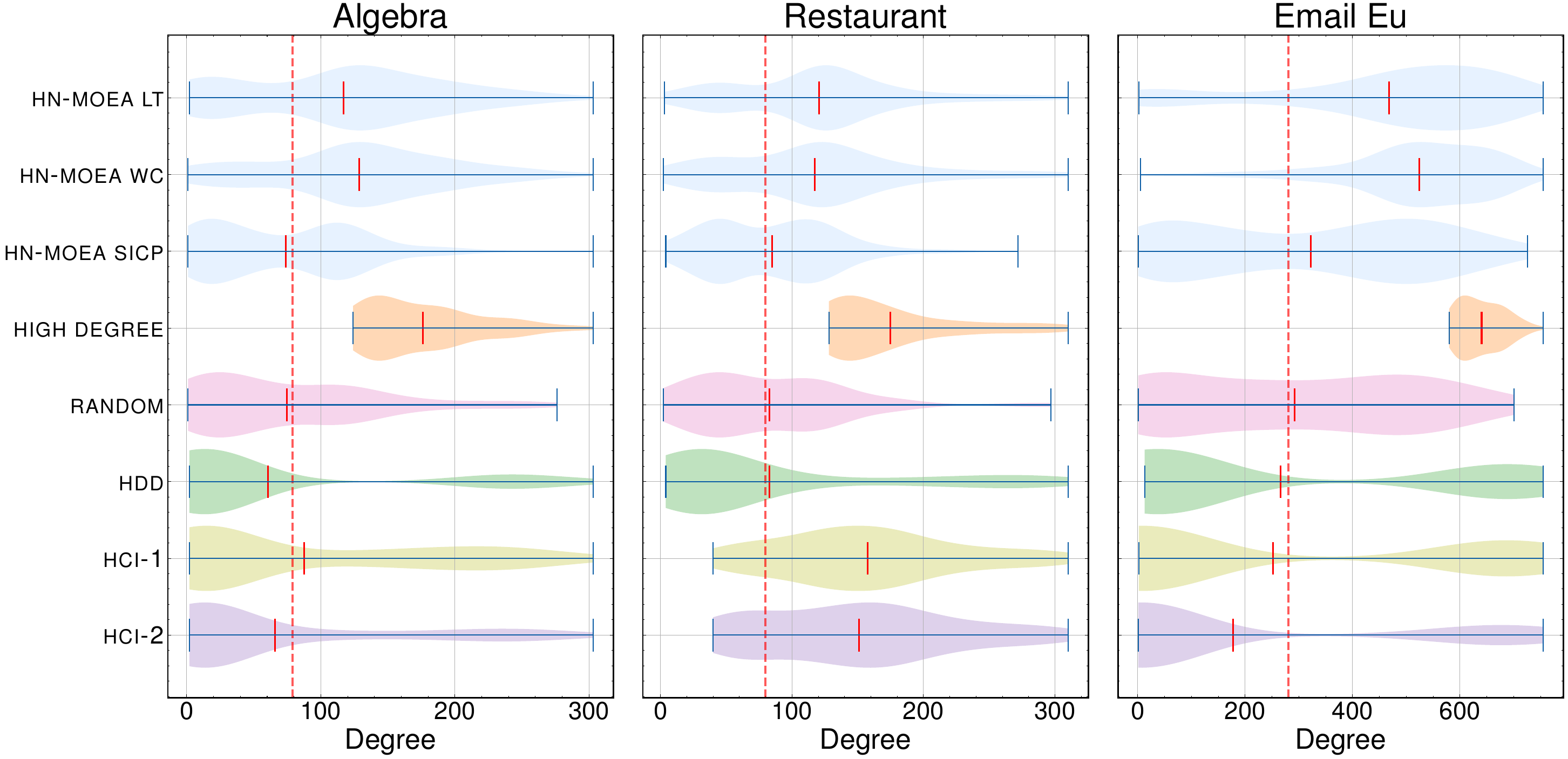}
    \caption{Violin plot w.r.t the Pareto Front of \method and baselines. The dotted red lines correspond to the avg. degree of the hypergraph. Red lines inside the violin plots correspond to the avg. degree of the solutions in the Pareto Front.}
    \label{fig:degree}
    \vspace{-0.5cm}
\end{figure}

\noindent \textbf{Performance in terms of solution diversity}
As introduced before, the bi-formulation of IM problem leads to avoiding having an incremental Pareto Front (as usual in single-objective formulation), providing more diversity in the solution seed sets of the final Pareto Front. Hence, we further enhance our analysis by comparing the characteristics of the nodes comprising the seed sets proposed by the IM algorithms under investigation.
In this regard, \Cref{fig:degree} highlights the \emph{topological diversity} in terms of the degree distribution of the nodes included in solutions found by each algorithm on three selected datasets (one per category). In the case of \method, we report the results for each of the three propagation models, while for the baselines the results do not depend on the propagation model as the seed set is built only based on the properties of the hypergraph. 
Upon examining the figure, we notice that \method incorporates nodes spanning a wide spectrum of degree values, which suggests that the EA benefits from the ability to explore the solution space without rigid adherence to specific greedy properties, as done in some of the compared heuristics. This usually leads to better results, although the effectiveness of node properties can vary depending on the peculiarities of the network at hand. In this regard, a node metric that works well as an indicator of a promising influence source in one dataset or region of a hypergraph may perform poorly in other contexts. For instance, high centrality does not invariably denote optimal influence propagation sources. As an example, bridge nodes linking distinct communities might have few neighbors, yet they are crucial for propagating influence.

In addition, 
\Cref{tab:diversity} reports the \emph{population diversity} and \emph{node diversity} within the solutions in the Pareto Front found by each algorithm. Note again that the solutions obtained by the baselines do not depend on the propagation model and as such they are all characterized by the same diversity.

\emph{Population diversity} refers to the extent to which individuals in the Pareto Front exhibit distinct genotypes. In our context, achieving a high level of population diversity relates to having minimal overlap between the seed sets of different individuals. Given a Pareto Front $\mathcal{P}$, where each individual $x_i \in \mathcal{P}$ is a seed set $x_i=\{v_1, v_2, \hdots, v_{|x|}\}$, the population diversity $D: \mathcal{P} \rightarrow \mathcal{R}$ is computed as:
\begin{equation}
    D(\mathcal{P})=1-\frac{1}{|\mathcal{P}|(|\mathcal{P}|-1)} \sum_{x_i \in \mathcal{P}} \sum_{x_j \in \mathcal{P}, i \neq j} \frac{|x_i \cap x_j|}
    {|x_i|}.
    \label{eq:d}
\end{equation} 

\emph{Node diversity}, instead, measures the percentage of unique nodes within the seed sets in the Pareto Front. A high node diversity indicates that the Pareto Front does not comprise nodes confined to a few regions of the network; rather, the proposed solutions represent a wide array of possible spread sources from various hypergraph locations. 
The node diversity $ND: \mathcal{P} \rightarrow \mathcal{R}$ is calculated as:

\begin{equation}
ND(\mathcal{P})=\frac{\left|\bigcup_{x_i \in \mathcal{P}}x_i\right|}{\sum_{x_i \in \mathcal{P}} |x_i|}
\label{eq:nd}
\end{equation}

As shown in the table, the inherent superior ability of \method to effectively explore the complex search space leads to a Pareto Front that not only encompasses nodes with heterogeneous topological features but also demonstrates higher population and node diversity compared to the solutions obtained by algorithms that construct seed sets by iteratively adding nodes. 

\begin{table}[ht!]
    \scriptsize
    \centering
    \caption{Population and node diversity achieved by the compared algorithms. Algorithms that incrementally construct seed sets by adding nodes to maximize a specific objective function inherently generate Pareto Fronts that exhibit the same values of node and population diversity. The boldface indicates the highest population and node diversity per dataset and propagation model.}
    \label{tab:diversity}
    \resizebox{.86\textwidth}{!}{
    \begin{tabular}{l l cccc}
    \toprule
     \textbf{Dataset} & \textbf{Diversity} & \textbf{\method - LT} & \textbf{\method- WC} & \textbf{\method- SICP} & \textbf{Baselines} \\
     \midrule
    \multirow{2}{*}{{Algebra}} & Population & $\mathbf{5.10\mathrm{e}{-01} \pm 4.00\mathrm{e}{-02}}$ & $3.43\mathrm{e}{-01} \pm 9.59\mathrm{e}{-03}$ & $3.89\mathrm{e}{-01} \pm 5.06\mathrm{e}{-03}$ & $2.50\mathrm{e}{-01}$ \\
    & Node  & $\mathbf{9.30\mathrm{e}{-02} \pm 2.66\mathrm{e}{-03}}$ & $4.39\mathrm{e}{-02} \pm 1.39\mathrm{e}{-03}$ & $4.79\mathrm{e}{-02} \pm 1.08\mathrm{e}{-03}$ & $2.00\mathrm{e}{-02}$ \\
    \midrule
    \multirow{2}{*}{{Geometry}} & Population & $\mathbf{5.54\mathrm{e}{-01} \pm 3.50\mathrm{e}{-02}}$ & $3.74\mathrm{e}{-01} \pm 9.62\mathrm{e}{-03}$ & $3.68\mathrm{e}{-01} \pm 1.10\mathrm{e}{-02}$ & $2.50\mathrm{e}{-01}$ \\
    & Node  & $\mathbf{9.78\mathrm{e}{-02} \pm 1.60\mathrm{e}{-02}}$ & $4.71\mathrm{e}{-02} \pm 3.13\mathrm{e}{-03}$ & $5.59\mathrm{e}{-02} \pm 4.78\mathrm{e}{-03}$ & $2.00\mathrm{e}{-02}$ \\
    \midrule
    \multirow{2}{*}{{MAG-10}} & Population & $3.50\mathrm{e}{-01} \pm 2.45\mathrm{e}{-02}$ & $3.36\mathrm{e}{-01} \pm 1.64\mathrm{e}{-02}$ & $\mathbf{3.99\mathrm{e}{-01} \pm 5.23\mathrm{e}{-03}}$ & $2.50\mathrm{e}{-01}$ \\
    & Node  & $\mathbf{9.41\mathrm{e}{-02} \pm 7.18\mathrm{e}{-03}}$ & $7.99\mathrm{e}{-02} \pm 4.81\mathrm{e}{-03}$ & $5.80\mathrm{e}{-02} \pm 6.23\mathrm{e}{-03}$ &$2.00\mathrm{e}{-02}$ \\
    \midrule
    \multirow{2}{*}{{Restaurant}} & Population & $\mathbf{5.16\mathrm{e}{-01} \pm 3.71\mathrm{e}{-02}}$ & $3.48\mathrm{e}{-01} \pm 2.53\mathrm{e}{-03}$ & $4.43\mathrm{e}{-01} \pm 1.85\mathrm{e}{-02}$ & $2.50\mathrm{e}{-01}$ \\
    & Node  & $\mathbf{9.11\mathrm{e}{-02} \pm 7.34\mathrm{e}{-03}}$ & $5.40\mathrm{e}{-02} \pm 3.86\mathrm{e}{-03}$ & $4.77\mathrm{e}{-02} \pm 3.33\mathrm{e}{-03}$ &$2.00\mathrm{e}{-02}$ \\
    \midrule
    \multirow{2}{*}{{Music}} & Population & $\mathbf{5.79\mathrm{e}{-01} \pm 3.93\mathrm{e}{-02}}$ & $3.61\mathrm{e}{-01} \pm 5.58\mathrm{e}{-03}$ & $4.52\mathrm{e}{-01} \pm 3.87\mathrm{e}{-02}$ & $2.50\mathrm{e}{-01}$ \\
    & Node  & $\mathbf{1.47\mathrm{e}{-01} \pm 1.90\mathrm{e}{-02}}$ & $5.61\mathrm{e}{-02} \pm 4.44\mathrm{e}{-03}$ & $5.04\mathrm{e}{-02} \pm 4.56\mathrm{e}{-03}$ & $2.00\mathrm{e}{-02}$ \\
    \midrule
    \multirow{2}{*}{{Bars}} & Population & $\mathbf{5.66\mathrm{e}{-01} \pm 2.80\mathrm{e}{-02}}$ & $3.85\mathrm{e}{-01} \pm 1.42\mathrm{e}{-02}$ & $5.20\mathrm{e}{-01} \pm 1.70\mathrm{e}{-02}$ & $2.50\mathrm{e}{-01}$ \\
    & Node  & $\mathbf{1.61\mathrm{e}{-01} \pm 4.58\mathrm{e}{-02}}$ & $6.90\mathrm{e}{-02} \pm 2.84\mathrm{e}{-03}$ & $6.02\mathrm{e}{-02} \pm 4.83\mathrm{e}{-03}$ & $2.00\mathrm{e}{-02}$ \\
    \midrule
    \multirow{2}{*}{{Email-eu}} & Population & $\mathbf{6.32\mathrm{e}{-01} \pm 6.14\mathrm{e}{-02}}$ & $4.41\mathrm{e}{-01} \pm 1.36\mathrm{e}{-02}$ & $4.75\mathrm{e}{-01} \pm 1.63\mathrm{e}{-02}$ & $2.50\mathrm{e}{-01}$ \\
    & Node  & $\mathbf{2.58\mathrm{e}{-01} \pm 3.60\mathrm{e}{-02}}$ & $5.21\mathrm{e}{-02} \pm 3.13\mathrm{e}{-03}$ & $5.89\mathrm{e}{-02} \pm 4.36\mathrm{e}{-03}$ & $2.00\mathrm{e}{-02}$ \\
    \midrule
    \multirow{2}{*}{{Email-enron}} & Population & $\mathbf{4.72\mathrm{e}{-01} \pm 1.90\mathrm{e}{-02}}$ & $3.11\mathrm{e}{-01} \pm 7.30\mathrm{e}{-03}$ & $4.57\mathrm{e}{-01} \pm 2.11\mathrm{e}{-02}$ & $2.50\mathrm{e}{-01}$ \\
    & Node  & $\mathbf{1.37\mathrm{e}{-01} \pm 1.78\mathrm{e}{-02}}$ & $7.83\mathrm{e}{-02} \pm 9.24\mathrm{e}{-03}$ & $5.59\mathrm{e}{-02} \pm 2.61\mathrm{e}{-03}$ & $2.00\mathrm{e}{-02}$ \\
    \midrule
    \multirow{2}{*}{{Email-w3c}} & Population & $\mathbf{4.30\mathrm{e}{-01} \pm 1.87\mathrm{e}{-02}}$ & $2.87\mathrm{e}{-01} \pm 1.06\mathrm{e}{-02}$ & $3.94\mathrm{e}{-01} \pm 2.10\mathrm{e}{-02}$ & $2.50\mathrm{e}{-01}$ \\
    & Node  & $\mathbf{1.25\mathrm{e}{-01} \pm 2.97\mathrm{e}{-02}}$ & $7.95\mathrm{e}{-02} \pm 4.50\mathrm{e}{-03}$ & $5.02\mathrm{e}{-02} \pm 1.71\mathrm{e}{-03}$ & $2.00\mathrm{e}{-02}$ \\
    \bottomrule
    \end{tabular}
    }
    \vspace{-0.2cm}
\end{table}

\section{Conclusions}
\label{sec:conclusion}
\vspace{-0.15cm}

In this paper, we presented \method, a multi-objective EA for IM in higher-order networks. To the best of our knowledge, this work marks the first attempt at employing EAs in this domain. Our method aims to minimize the seed set size $|S|$ while maximizing the expected influence, and includes smart initialization and hypergraph-aware mutations to improve convergence and performance.

The method has been evaluated on nine different real-world datasets characterized by heterogeneous properties, with three different propagation models. Our experimental analysis demonstrates that \method overall outperforms current state-of-the-art algorithms for IM in higher-order networks. In line with previous findings~\cite{moea-for-im-in-social-networks, improving-multi-objective,cunegatti2022large}, these results confirm that EAs are particularly well-suited for solving discrete optimization problems of this nature.

One notable advantage of \method lies in its flexibility to maximize influence across various propagation models. Moreover, in contrast to other heuristic methods, \method explores the solution space without any bias towards specific node metrics and greedy properties. As a result, the evolutionary process better explores the search space that characterizes the IM problem. Because of the bi-objective formulation, the resulting population also exhibits higher individual diversity, with candidate solutions comprising nodes of varied properties.

Future research could focus on many-objective IM, as recently done in~\cite{cunegatti2024many} in the context of standard graphs. Moreover, in this study, we employed hypergraph-aware mutations chosen at random. However, the selection of the optimal mutation operator may vary depending on the characteristics of the dataset and the evolutionary stage. Hence, future research could investigate the use of adaptive mutation operators to select the most suitable mutation operator based on feedback from the search process.

\bibliographystyle{splncs04}
\bibliography{refs}


\clearpage

\end{document}